\documentclass[twocolumn]{revtex4}

\usepackage{graphicx}
\usepackage{dcolumn}
\usepackage{bm}

\begin{document}

\title{Comment on`` Angular dependence of Dicke-narrowed electromagnetically induced transparency resonances''}

\author{S. M. Iftiquar}
\affiliation{Department of Physics, Indian Institute of Science, Bangalore - 560012, INDIA}

\date{\today}

\begin{abstract}
We demonstrate that the experimental results reported by M. Shuker, O. Firstenberg, R. Pugatch, A. Ben-Kish,1 A. Ron, and N. Davidson, Phys. Rev. A {\bf 76}, 023813 (2007) as Dicke narrowing of electromagnetically induced transparency, does not match  with theory.  
\end{abstract}


\maketitle

In a recent article [1] Shuker et al. claim to observe electromagnetically induced transparency (EIT) at various angles between pump and probe and conclude a good quantitative agreement between measurement and theoretical model. 

We found the experimental measurement and  theoretical model are not matching with each other. The energy level scheme adopted in this experiment, as shown in figure 1(a), is prone to give EIT width much greater than $10$ KHz, because of Clebsch Gordan coefficients of the concerned energy levels and decay rates $\gamma$ of $F^{\prime}=1, m_{F^{\prime}}=+1 \rightarrow F=2, m_{F^{\prime}}=+1$ channel, where $\gamma$ is more than 10 KHz. This $\gamma$ should come within the expression $\Gamma_{12}+(2 \pi L/\lambda)\Gamma_{D} \theta^{2}$. Furthermore the combination of hyperfine levels adopted in this report is different from that generally adopted for EIT [2]. Thus the experiment is very unlikely to give such a narrow EIT signal.

In the experimental part, the laser is locked to $F=2 \rightarrow F^{\prime}=1$ transition and at the sametime there is acousto-optic modulator (AOM) along the beam path, which deviates the laser line from the $F=2 \rightarrow F^{\prime}=1$ transition frequency. So the actual laser frequency with which the experiments were performed may be different from that shown in figure 1(a). Usually the AOMs respond at about 80 MHz radio frequency although AOMs of frequency above 500 MHz and below 60 MHz may be available. As the authors have not specified the AOM frequency and laser frequency shift, so the experiment is most likely to have been carried out connecting $F^{\prime}=2$ which is a better energy level in combination with $F=2$.

Although the figure 5 in [1] is very interesting yet it seems misleading at a first glance that while EIT happens outside the EIT zone absorption is maximum. We do not see such total suppression of probe transmission even for resonant probe beam in a Rb vapour cell or any such similar condition; fundamentally this violates Beer Lambert's law of exponential decay in transmitted beam intensity.

To conclude, it is most likely that the energy level diagram is not right and the narrow resonance signal might come from some other physical process or the specific instruments used. There is also high possibility that images of figure 5 does not fully signify roles of angle dependant EIT on transmitted beam intensity.




\end{document}